\documentclass[preprint,aps,prl,overcite]{revtex4}
\usepackage{epsfig}
\usepackage{amsmath}

\newcommand\kessence{\textit{k}-essence}
\newcommand\CMB{\textsc{cmb}}
\newcommand\CMBFAST{\textsc{cmbfast}}
\newcommand\CDM{\textsc{cdm}}

\begin{document}

\title{Measuring the Speed of Sound of Quintessence}
\author{Joel K. Erickson$^1$, R.R. Caldwell$^2$, Paul J. Steinhardt$^1$, C.
Armendariz-Picon$^3$, V. Mukhanov$^4$}
\date{July 2001}
\affiliation{$^1$Department of Physics, Princeton University, Princeton, NJ 08544 \\
$^2$Department of Physics \& Astronomy, Dartmouth College Hanover, NH 03755 \\
$^3$
Department of Astronomy and Astrophysics, 
Enrico Fermi Institute,
University of Chicago, Chicago, IL 60637 \\
$^4$Ludwig Maximilians Universit\"{a}t, Sektion Physik, 80333 M\"{u}nchen,
Germany }

\begin{abstract}
Quintessence, a time-varying energy component that may account for the
accelerated expansion of the universe, can be characterized by its equation
of state and sound speed. In this paper, we show that if the quintessence
density is at least one percent of the critical density at the 
surface of last
scattering the cosmic microwave background anisotropy can distinguish
between models whose sound speed is near the speed of light versus near
zero, which could be useful in distinguishing competing candidates for dark
energy.
\end{abstract}

\maketitle


Recent evidence suggests that most of the energy density of the universe
consists of a dark energy component with negative pressure. Two candidates
are a cosmological constant (or vacuum density) and 
quintessence,\cite{Cal98} a 
time-varying energy component with negative pressure.  
A common example of quintessence is a  scalar
field slowly rolling down its potential.  
The scalar field may be regarded as real, or simply as a device
for modeling more general  cosmic fluids with negative pressure.
A cosmological constant and quintessence
can, in principle, be distinguished
by the equation of state, $w$, the ratio of the pressure $p$ to the energy
density $\rho $. A cosmological constant  always has $w=-1$ whereas a scalar
field generally has $w$ different from unity and time-varying. Through
measurements of supernovae, large-scale structure and the cosmic microwave
background (\textsc{cmb}) anisotropy, the equation of state may be
determined accurately enough in the next few years to find out whether $w$
is actually different from $-1$ or not.

For a  scalar field with a negative effective pressure and slowly
varying energy density, we can further  define two cases.
The first consists of  standard models of
quintessence\cite{Cal98,quint}
in which   the scalar field $\phi$ has
a canonical kinetic term,  $X=\frac{1}{2}(\partial _{\mu }\phi )^{2}$. 
A second 
 category  consists of models in which the kinetic energy
is not canonical.
Prominent examples are    
 \textit{k}-essence\ models, which are designed to
address the issue of why cosmic acceleration has recently begun.\cite{Arm00}
The
equation of state in \textit{k}-essence\
models is positive until the onset of
matter-domination triggers a change to negative pressure. The models rely on
dynamical attractor behavior due to the non-canonical kinetic energy
density. The Lagrangian density  is a rather generic
function of $X$: $\mathcal{L}_{\phi }(X,\phi )=\widetilde{p}%
(X)/\phi ^{2}.$ 

One difference between standard quintessence
and \textit{k}-essence models 
is the time-evolution of the equation of state.
 The equation of state for a \textit{k}-essence component approaches $-1$
 soon after the onset of matter-domination and then increases towards a less
 negative value in the present epoch 
 as the component begins to dominate the energy density.
In standard 
quintessence models,  the equation of state  is generically  monotonically
decreasing and approaching $-1$ today.
This feature was discussed in detail in Ref.~3. 

In this paper, we focus on  a second physical property -- the speed
of sound  -- which also distinguishes 
standard quintessence  from \textit{k}-essence and, more generally,
from other cosmic fluids described by a non-canonical 
kinetic energy density.  
Let us consider a
scalar field $\phi $ with a general Lagrangian density $\mathcal{L}_{\phi
}(X,\phi )$. The stress-energy tensor in this case can be expressed in the
form  of an ideal fluid
\begin{equation}
T^{\mu \nu }=\left( \rho +p\right) u^{\mu }u^{\nu }-pg^{\mu \nu },
\end{equation}%
where the pressure is $p=\mathcal{L}_{\phi }$, the energy density is $\rho =2%
\mathcal{L}_{\phi ,X}X-\mathcal{L}_{\phi }$ and $,X$ denotes the  partial
derivative with respect to $X.$ From this, we find 
\begin{equation}
w=\frac{\mathcal{L}_{\phi }}{2\mathcal{L}_{\phi ,X}X-\mathcal{L}_{\phi }}.
\label{eqofstate}
\end{equation}%
The effective speed of sound entering the equations which describe the
evolution of small perturbations is 
\begin{equation}
c_{s}^{2}=\frac{p_{,X}}{\rho _{,X}}=\frac{\mathcal{L}_{\phi ,X}}{\mathcal{L}%
_{\phi ,X}+2\mathcal{L}_{\phi ,XX}X}.
\end{equation}%
Thus, for the standard quintessence models
with canonical scalar fields, $\mathcal{L}=X-V(\phi )$,
the equation of state is
\begin{equation*}
w=\frac{X-V}{X+V}
\end{equation*}%
and the speed of sound is always equal to one: $c_{s}^{2}=1.$ However, if $%
\mathcal{L}_{\phi ,XX}\neq 0,$ 
then $c_{s}^{2}\neq 1$. (In fact, even $c_{s}^{2}>1$ is possible.
Physically, this means that perturbations of the background scalar field can
travel faster than light as measured in the preferred frame where the
background field is homogeneous.  For a time dependent
background field, this is not a Lorentz invariant state. 
There is no violation of
causality. The underlying theory is manifestly Lorentz
invariant and it is not possible to transmit information faster than light
along \textit{arbitrary} space-like directions or create closed time-like
curves.)

This paper  investigates how the variable  speed of sound
influences the fluctuations of the cosmic
microwave background compared to the case of standard
quintessence where $c_{s}^{2}=1$.  In general, the effect is small,
but we show that it is detectable in cases
like 
 \textit{k}-essence models in which 
the speed of sound is nearly zero
during most of the period between last scattering and the present epoch.
This behavior 
produces the greatest difference from standard quintessence.\cite{Hu}
We first compare models with exactly the same equation of state as a
function of redshift, $w(z)$, but different sound speed. We find that models
with near-zero sound speed today (such as \textit{k}-essence\ models) are
distinguishable from models with $c_{s}=1$ based on measurements of the 
\textsc{cmb}\ power spectrum, provided the quintessence energy density is
greater than a few percent of the critical density at last scattering. The
density requirement, which is satisfied by typical \textit{k}-essence\
models, for example, is needed so that the sound speed has a measurable
effect on the acoustic oscillation peaks of the \textsc{cmb}, which are
sensitive to conditions at the last scattering surface. 
Similar results can be obtained for more general forms 
of dark energy.\cite{Hu,Chiba}
We then consider
whether the effect can be mimicked by varying other cosmic parameters or by
introducing a time-varying equation-of-state. To perform the studies, we
introduce a spline technique that is useful in exploring models with
time-varying $w$. Our conclusion is that the sound speed effect is
distinguishable from all other standard parameter effects. Hence, the 
\textsc{cmb}\ can provide a useful constraint on the sound speed of dark
energy.



\begin{figure}
\begin{center}
\epsfig{file=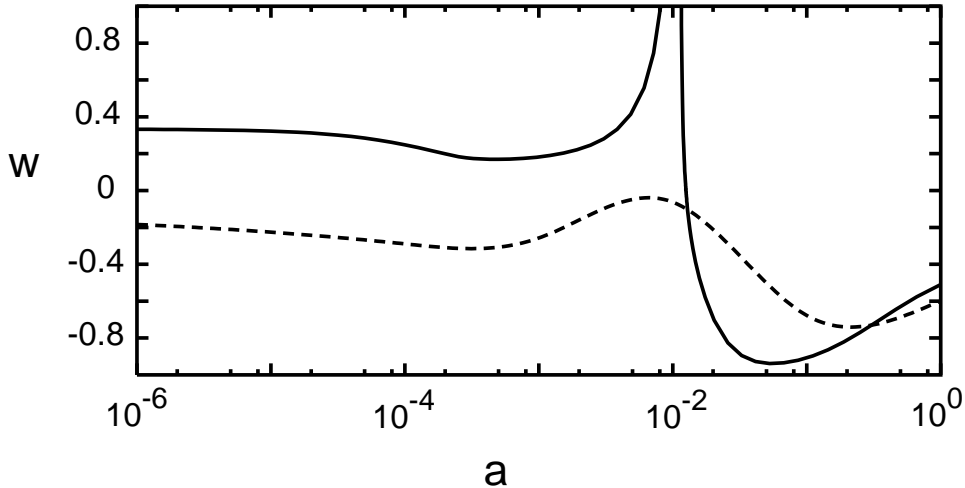,width=5in}
\end{center}
\caption{
The equation of state for the fiducial model (solid line) and the
closest spline fit (dashed line) quintessence model. The 
spline fit is chosen so that the CMB spectrum matches the fiducial
model to within the cosmic variance limit. However, the spline itself
is smoother than the actual $w(a)$. As discussed in the text,
there is a large degeneracy in
the spline parameters, so the spline need not mimic the fiducial equation 
of state very closely to obtain a good match to the CMB spectrum.}
\label{fig:eos}
\end{figure}

\begin{figure}
\begin{center}
\epsfig{file=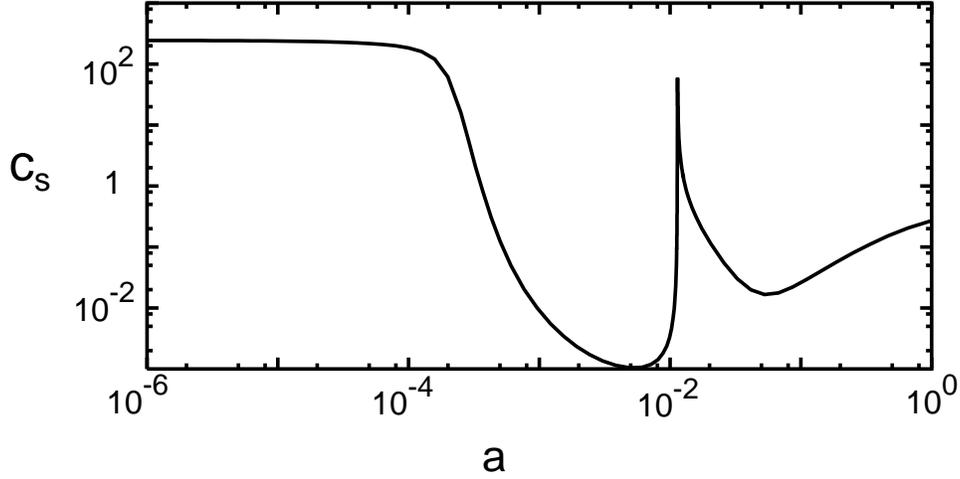,width=5in}
\end{center}
\caption{The speed of sound as a function of the scale factor $a$ 
($a_{\text{today}}=1$). Note that at $c_s^2\ll 1$ near the last 
scattering surface and at the present epoch.}
\label{fig:cs}
\end{figure}

\begin{figure}
\begin{center}
\epsfig{file=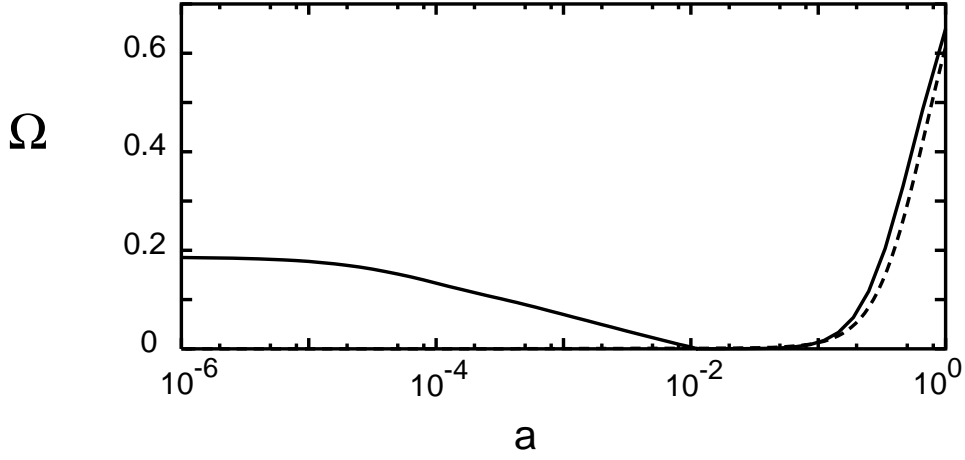,width=5in}
\end{center}
\caption{$\Omega_{\mathrm{\kessence}}$ and $\Omega_{\mathrm{Q}}$ as a
function of $z$ for the \kessence\ (solid line) and best-fit spline
models (dashed line), respectively. Note that $\Omega_Q$ falls off to
zero at large redshifts, whereas $\Omega_{\kessence}$ approaches a
finite value; $w$ approaches $1/3$ for this model at large redshifts.}
\label{fig:omega}
\end{figure}

In the case of \kessence\, the Lagrangian generally has the form
$\mathcal{L}_\phi=\widetilde{p}(X)/\phi^2$, 
where $\widetilde{p}(X)$ is generally some
function with $\widetilde{p}_{,XX}\ne 0$. We have 
chosen a specific form for $\widetilde{p}$
for our fiducial model. The equation of state and sound speed for this
model are shown in Figs~\ref{fig:eos} and \ref{fig:cs} respectively; 
they are expressed as
functions of the scale factor $a$ ($a=1$ today) by integrating the
equations of motion. To completely specify the model, we fix the
cosmological parameters today to typical values: $\Omega_b=0.05$,
$\Omega_{\text{CDM}}=0.3$, $\Omega_Q=0.65$ and $h_0=0.5$ (where $h_0$
is the Hubble parameter in units of
$100\mathrm{\,km\,sec^{-1}\,Mpc^{-1}}$). The energy density as a
fraction of the critical density, $\Omega_Q$ (\kessence), is shown in
Fig.~\ref{fig:omega}.  While $c_s^2>1$ for large redshifts, this is
not an important feature of the model: $\Omega_Q$ 
 is reasonably small whenever $c_s^2>1$, so the
value of $c_s$ at these times has negligible effect on the \CMB.  (We
have verified this by rerunning the calculation after artificially
truncating the speed of sound at $c_s\le 1$ and comparing the
\CMB\
power spectra. 
Also, we should note that, with a slightly different 
parameterization, we can obtain a model in which $c_s <1$ which 
at early times and which 
has the same behavior at late times.)
The most important feature of the fiducial
model is that $c_s^2\ll 1$ whenever \kessence\ contributes
significantly to the energy density of the universe.

The \CMB\ power spectra for our models are computed by modifying the
standard \CMBFAST\ code.\cite{Sel96,Cal98}  Simply put, the effect
of the speed of sound on the \CMB\ perturbation equations is such that
for $c_s^2\ll 1$, \kessence\ will collapse via gravitational
instability into cold dark matter (\CDM) potentials, whereas in the
superluminal case, $c_s^2>1$, the growth of density perturbations is
suppressed.  For those familiar with the code, the modifications are 
straightforward.
The perturbed line element is
\begin{equation}
ds^2 = a^2(\eta)[ d\eta^2 - (\delta_{i j} + h_{i j})dx^i dx^j]
\end{equation}
where $\delta_{ij}$ is the unperturbed spatial metric, and $h_{ij}$ is the
metric perturbation. We shall use
 $h$ to represent
 the trace of the spatial metric perturbation.
The effect we are examining is due
to the perturbations  to the $k$-essence stress-energy 
in the synchronous gauge for a mode with wavenumber $k$
\begin{eqnarray}
\delta\rho &=& -2 \rho \frac{\delta\phi}{\phi}
-(\rho+p)\frac{\delta y }{ y} c_s^{-2} 
\cr\cr
\delta p &=& - 2 p \frac{\delta\phi}{ \phi}
-(\rho+p)\frac{\delta y }{ y} 
\cr\cr
\theta &=& \frac{1}{\sqrt{2}} k^2 y \delta\phi
\end{eqnarray}
where $y \equiv 1/\sqrt{X}$ and
$\theta$ is the divergence of the fluid velocity.
The density contrast, $\delta 
\equiv \delta \rho/\rho$, obeys the equation
\begin{equation}
\dot\delta = -(1+w)\left(\theta + \frac{1}{2} \dot h\right) 
- 3 {\cal H}  \left( \frac{\delta p}{ \delta \rho} - w \right) \delta,
\label{deltaeqn}
\end{equation}
where the derivative is with respect to conformal time
and ${\cal H} \equiv \dot{a}/{a}$.
Notice that the quantity $\delta p / \delta\rho$ appears, which is not
to be confused with $c_s^2$. In fact
\begin{equation}
\delta p = c_s^2 \delta\rho + \frac{\theta \rho }{ k^{2}}  
\left[ 3 {\cal H}  (1+w)(c_s^2 - w) + \dot w \right]
\label{deltapeqn}
\end{equation}
which leads to a simplified evolution equation for the velocity gradient
\begin{equation}
\dot\theta = (3 c_s^2 - 1) {\cal H} \theta   + c_s^2 k^2 \delta / (1 + w).
\label{deltateqn}
\end{equation}
The kinetic quintessence is distinct from scalar field quintessence, for which
$c_s^2=1$ in equations  (\ref{deltapeqn}) and  (\ref{deltateqn}).
Linearized perturbations in $k$-essence can propagate  non-relativistically, with
$c_s^2 \ll 1$. We can see in equation (\ref{deltateqn}) that a small sound
speed will cause the velocity gradient to decay; with the conventional gauge
choice that $\theta_{cdm}=0$, the inhomogeneities in the  $k$-essence will
describe a fluid which is comoving with the cold  dark matter. From equation
(\ref{deltapeqn}), we see that the second term on the RHS
will be negligible even on scales approaching the horizon. The overall
effect is that the pressure fluctuations $\delta p$ are too weak to 
prevent
$k$-essence collapse via gravitational instability into the 
CDM gravitational
potentials.

The  \CMBFAST\ 
code takes $w(a)$ and $c_s(a)$ as inputs, so it is
possible to manually adjust these functions to have any values
(including, of course, $c_s = 1$). Once we have computed the \CMB\
anisotropy for two models, they can be compared by computing their
likelihood difference, the probability that they could be confused due
to the cosmic variance in local measurements of the \CMB.
Given models $A$ and
$B$, the negative log-likelihood, $-\log L$, is derived in
\cite{Hue99}:
\[
-\log L=\sum_l(l+{\textstyle\frac{1}{2}})\Bigl(1-\frac{C_{lA}}{C_{lB}}+\log\frac{C_{lA}}{C_{lB}}\Bigr).
\]
The condition $-\log L>6$ corresponds to distinguishability at the
$3\sigma$ level or greater. The relative normalization of the spectra
is chosen so as to minimize the likelihood difference, including $l$
up to $1500$. 

\begin{figure}
\begin{center}
\epsfig{file=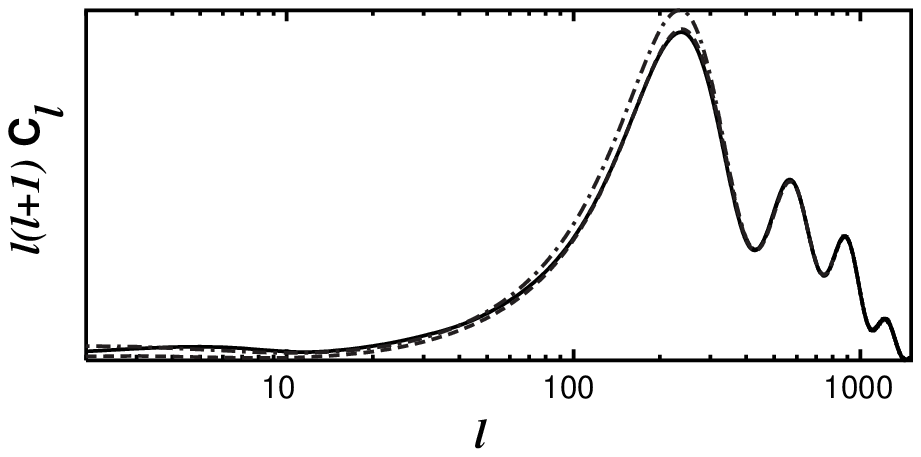,width=5in}
\epsfig{file=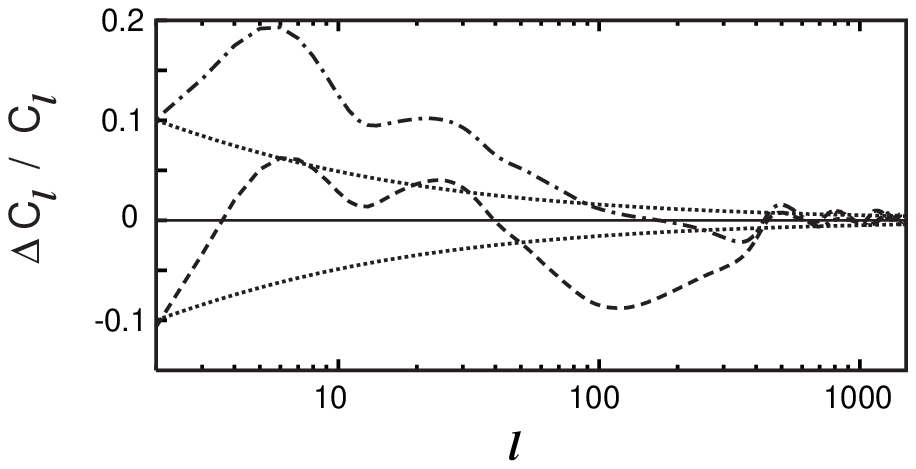,width=5in}
\end{center}
\caption{The \textsc{CMB} power spectrum for \kessence\ (solid line),
the model with $c_s=1$ (dot-dashed line, $-\log L=127$) and the best-fit
spline model (dashed, $-\log L=28$). Models that are
distinguishable can still fit quite closely. The lower diagram shows
$\Delta C_l/C_l$ (relative to the fiducial model) and the cosmic
variance envelope (with no particular normalization).}
\label{fig:cmb}
\end{figure}

\begin{figure}
\begin{center}
\epsfig{file=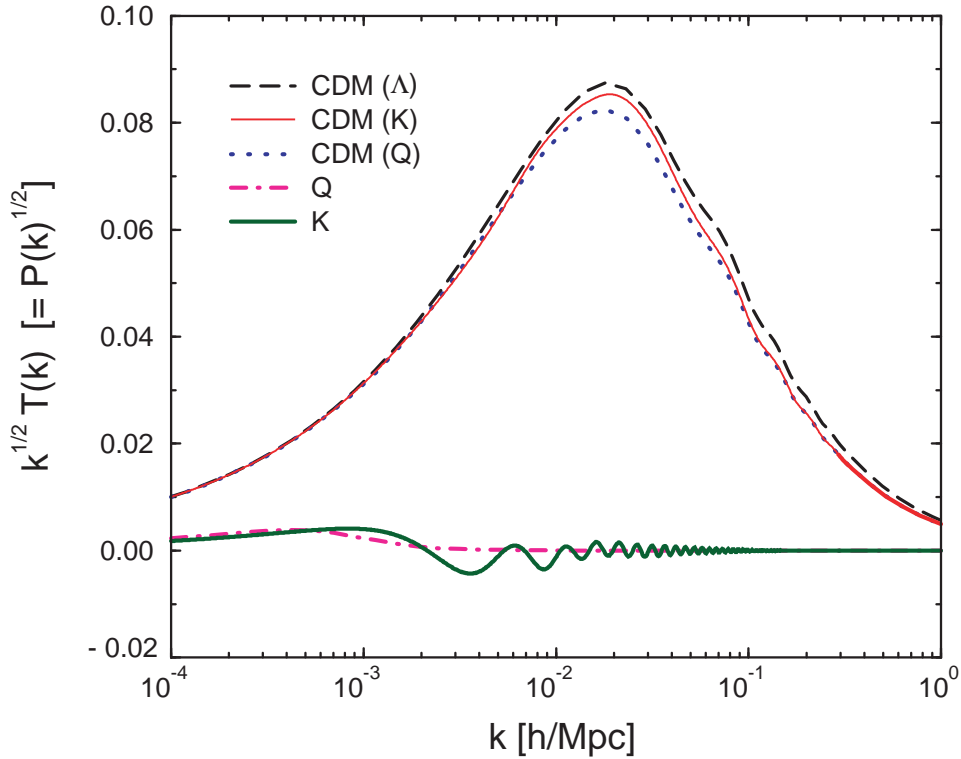,width=5in}
\end{center}
\caption{ Comparison of the power spectra for dark matter (CDM)
and dark energy ($Q$ or $K$)  for models with dark energy in the form of
cosmological constant ($\Lambda$), quintessence ($Q$) and $k$-essence
($K$). All models have $\Omega_{CDM}=0.3$, $\Omega_b h^2 =0.02$,
and $h=0.65$.  For models with cosmological constant, there is no perturbed
dark energy component.  Note
the distinctive oscillations of the $k$-essence component
associated with having $c_s \ll 1$.
}
\label{fig:pk}
\end{figure}

The power spectrum for our fiducial model is given by the solid line
in Fig.~\ref{fig:cmb}. We can see the effect of the unusual speed of
sound on this model by computing a new spectrum, which has the same
equation of state and cosmological parameters, but whose speed of
sound has been set equal to $1$ for all $a$. This corresponds to a
quintessence field -- with canonical kinetic energy -- rolling down a
potential. The power spectrum for the  model is shown in
Fig.~\ref{fig:cmb}. The two models have log likelihood
difference $-\log L=127$: they are easily distinguishable. The speed
of sound has a significant effect on the \CMB\ anisotropy.
Figure~\ref{fig:pk} compares the dark matter and dark energy 
contributions for quintessence and $k$-essence models.  The 
small sound speed 
 results in distinctive oscillations in the  case of $k$-essence.

Can the effect of the sound speed be distinguished from that of other
cosmological parameters? There is already a large degeneracy
\cite{Hue99} in these parameters, so it would not have been too
surprising if allowing a variable speed of sound merely expanded the
pre-existing degeneracy. This problem is addressed by considering, as
above, quintessence models which have $c_s^2  = 1$, but allowing
the values of $\Omega_b$, $\Omega_{\text{CDM}}$, $\Omega_Q$
(quintessence) and $h_0$ to vary (subject to the flatness condition
$\Omega_b+\Omega_{\text{CDM}}+\Omega_Q=1$). For the comparison models,
the equation of state, $w$, is taken to be constant as a function of
the scale factor $a$, but is allowed to vary from model to
model. Minimizing the log-likelihood over these parameters, the best
fit gives $-\log L=32$, with parameters $\Omega_b=0.05$,
$\Omega_{\text{CDM}}=0.35$, $\Omega_Q=0.60$, $h_0=0.48$ and
$w=-0.78$. This fit was found using well known minimization schemes.
\cite{Pre93} It seems likely that $-\log L=32$ is the best that can
be done for this class of models, as the result is quite insensitive
to the parameter values with which the mimization is started. Hence,
the speed of sound \textit{is} distinguishable.

Thus far, our fiducial model has been compared with two kinds of
$c_s =1$ models: one with all other parameters, including $w(a)$,
identical and one with constant $w$, with the parameters (including
$w$) adjusted so as to minimize the likelihood difference. In both
cases the fiducial model was easily distinguishable. We next test the
possibility that some other form of quintessence, with general $w(a)$
and canonical kinetic energy, can reproduce the \CMB\ anisotropy of
the time-varying $c_s$ model. For this purpose, we consider models
with an equation of state given by a cubic spline.\cite{Pre93}  That
is, we introduce six new parameters into our model: the values of $w$
at $a=10^{-4}$, $10^{-3}$, $10^{-2}$, $10^{-1}$ and $w$ at the two
extremes of $a$, which lie at $a=10^{-14}$ and $3.8$. (Introducing
more spline points has a negligible effect on our results.)
The equation of
state at other values of $a$ is then given by a piecewise cubic (in
$\log a$) function whose coefficients are chosen so that it passes
through the points selected above and has a continuous second
derivative. We still allow $\Omega_b$, $\Omega_{\text{CDM}}$,
$\Omega_Q$ and $h_0$ to vary (again enforcing flatness), and now
allow, for completeness, the spectral tilt $n_s$ to vary as well. The
model therefore has a total of ten free parameters. The minimum
log-likelihood difference found was $-\log L=28$, which is not a significant
improvement. This model has $\Omega_b=0.05$,
$\Omega_{\text{CDM}}=0.34$, $\Omega_Q=0.61$, $h_0=0.47$,
$n_s=1.02$. The equation of state is shown in Fig.~\ref{fig:eos}
 and the \CMB\ power spectrum is shown in
Fig.~\ref{fig:cmb} (dotted line).

The spline technique always produces a very smooth looking equation of
state, compared to the rapidly varying equation of state
(particularly near $a=0.01$) given by the actual \kessence\ model. 
To see that this does not affect the analysis -- that the
spline equation of state has sufficient freedom to closely mimic that
for the fiducial model -- we compare two models, one with $c_s=1$ and
one with an identical speed of sound to the \kessence\ model. The
equations of state of these models are allowed to vary using the
spline technique, but the cosmological parameters are fixed to be the
same as for the fiducial model. The minimum $-\log L$ for the $c_s=1$
model is $70$, whereas the log-likelihood difference for the model with
the fiducial sound speed is much less than one. Thus, the 
spline technique appears
to do a very good job of modeling the relevant details of the
equation of state.

In sum, we have seen from our example
that it is possible to distinguish models with
$c_s=1$ (e.g.\ models with canonical kinetic energy density) from
models with $c_s^2\ll 1$, as occurs in models with non-canonical
kinetic energy density, such as \kessence. Our studies 
show that distinction depends on $c_s \ll 1 $ and 
$\Omega_Q$ being at least a few percent at the last scattering surface
 so that the small value of the speed of sound 
affects the acoustic peaks, which can be precisely measured, as 
well as the large-angular scale anisotropy.
If $\Omega_Q \ll 1$ at the last scattering surface, then the speed
of sound only affects the large angular scale anisotropy and is 
difficult to distinguish because of the large cosmic variance
at those scales.  

As a technical device for the study,  we have introduced a cubic
spline method for studying varying a general equation of state $w(a)$
with a finite number of fitting quantities. The same technique can be
extended to include general time-varying $c_s$. In this way,
near-future observations of the \CMB\ may be used to constrain general
models of quintessence without introducing strong priors into its
nature.

\medskip

This work was supported by 
NSERC of Canada (JKE), NSF grant PHY-0099543 (RRC),
SFB375 der Deutschen Forschungsgemeinschaft (VM),
and 
Department of Energy grants 
DE-FG02-90ER4056 (CAP) and DE-FG02-91ER40671 (PJS).

\newpage

\end{document}